\documentclass[aps,prl,superscriptaddress,twocolumn,floatfix]{revtex4}

\usepackage{amsfonts,amssymb}
\usepackage{graphicx,pstricks}

\begin{document}


\title{ Rate determining factors in   protein model structures
}


\author{Pierpaolo Bruscolini}
\email{pier@unizar.es}
\affiliation{Instituto de Biocomputaci\'on y F\'isica de Sistemas Complejos (BIFI),
Universidad de Zaragoza, c.\ Corona de
  Arag\'on 42, 50009 Zaragoza, Spain}

\author{Alessandro Pelizzola}
\email{alessandro.pelizzola@polito.it}
\affiliation{Dipartimento di Fisica and CNISM, Politecnico di Torino,
  c. Duca degli Abruzzi 24, 10129 Torino, Italy}
\affiliation{INFN, Sezione di Torino}

\author{Marco Zamparo}
\email{marco.zamparo@polito.it}
\affiliation{Dipartimento di Fisica and CNISM, Politecnico di Torino,
  c. Duca degli Abruzzi 24, 10129 Torino, Italy}


\begin{abstract}
Previous research has shown a strong correlation of protein folding rates to
the native state geometry, yet a complete
explanation for this dependence is still lacking.
Here we study the rate-geometry relationship with a simple
statistical physics model, and
focus on two classes of model geometries, representing ideal parallel and antiparallel structures. 
We find that the logarithm of the rate shows
an almost perfect linear correlation with the "absolute
contact order", but the slope depends on the particular
class considered. We discuss these findings in the light of experimental results.
\end{abstract}


\maketitle


\label{sec:intro}
In the last decade many results have been published on the relationship between kinetic and structural features of protein folding, after the seminal work by  Plaxco, Simons and Baker
\cite{Plaxco} revealed a simple linear relationship between the logarithm of the folding rate and the relative contact order (CO) for a set of two-state
proteins.
CO is defined as the sum of the sequence distances $\vert j-i\vert $ of any residue pair $(i,j)$ in contact in the native structure (with a "contact" being defined according to a cutoff distance), divided by the total number of contacts and number of residues.
A similar good correlation was obtained by
Jackson \cite{Jackson}, considering the absolute contact order (ACO, equal to CO times the number of residues)
and using a slightly extended protein set. Galzitskaya {\it et al.} in \cite{GalFinkProt2003} showed that length, rather than CO, is relevant in three-state proteins. The role of the length is also considered in \cite{Thir95,Cieplak,Thir2004,MunozJACS}, 
while  the logarithm of the relative contact order, together with CO and ACO, was considered in
\cite{Grantcharova}, and a combination of CO and length in
\cite{Koga}.
The rates have been  also related to linear combinations of the length and its logarithm \cite{Makarov}, long range order (LRO), total
contact distance \cite{Zhou}, fraction of short range contacts \cite{Miller}, cliquishness \cite{Micheletti}, effective contact
order \cite{Dixit}. Plotkin and coworkers \cite{Plotkin} discussed the correlation with
heterogeneities in contact distance and energy, while the absolute contact order,
and its length dependence, was reconsidered in \cite{Ivankov}.
On the basis of the cited (and certainly not exhaustive) set of
investigations it is difficult to reach a definite conclusion about
which measures of native state properties are most relevant to
determine the folding rate. 
Indeed, while experimental rates show a clear correlation with the structural parameters proposed in the above literature,  they are always distributed with a relevant spread around the theoretical curve, that remains to be accounted for.  On- and off-lattice theoretical modelling \cite{jewpandeplaxco2003, kayachan2003} suggest that adding cooperativity and/or local preferential conformations to G\=o like models improves the correlation of the rates with CO, also inducing spread in the rate distribution.  But also the high heterogeneity of the distribution of the sequence-distance of the contacts in protein structures could be responsible for the rate spread \cite{Plotkin}.  
For the above reasons, it is very 
important to try to
address this issue with model structures, in the framework of a simple 
model with few parameters.
We focus on two classes of 
structures, the ideal analogues of real parallel  and antiparallel secondary structures ($\alpha$--helices and $\beta$--sheets): our aim is to find a set of rules 
that may be later used to rationalize the relationship between real protein geometries and rates.

We resort to  the Wako--Sait\^{o}--Mu\~noz--Eaton (WSME) model,
which allows an exact solution for the equilibrium
thermodynamics and a very accurate semi--analytical approach to the
kinetics.  
At difference with other native-centric models, WSME is intrinsically cooperative and accounts for the local preferences of the main-chain. 
The model has already been shown
to give good results in the determination of rates of real proteins
\cite{ME3,HenryEaton}, their temperature dependence \cite{ZP-PRL}, the
effect of mutations \cite{ZP-Prep} and mechanical unfolding rates
\cite{IPZ}. 
The WSME model 
was introduced in the end of the
'70s \cite{WS1,WS2} and then forgotten
for roughly 20 years, 
when it was indepentently reproposed by Mu\~noz, Eaton and
coworkers \cite{ME1,ME2,ME3,HenryEaton}. Several studies followed
\cite{Amos1,Amos2,ItohSasai1,ItohSasai2,ItohSasai3,AbeWako,BP,P,ZP-PRL,ZP-JSTAT,IPZ}
also with applications in a very different subfield
of physics \cite{TD1,TD2,TD3}.
WSME is a G\={o}--like model \cite{Go}, i.e. it is "native-centric", relying on the knowledge of
the native state of a protein to describe its equilibrium and kinetics. 
Its binary degrees of freedom 
are related to the values of the dihedral angles 
at the peptide bonds \cite{ME2}, classified 
into just two states: ordered (native) and disordered
(unfolded). Since the latter state allows a much larger number of
microscopic realizations than the former, an entropic cost is
given to the ordering of a peptide bond.

The model is described by the 
effective free energy:
\begin{equation}
H = \sum_{i=1}^{N-1}\sum_{j=i+1}^{N}
\epsilon_{i,j} \Delta_{i,j} \prod_{k=i}^{j} m_k
- R T \sum_{k=1}^{N} q_k (1-m_k) ,
\label{Hamiltonian}
\end{equation}
where $N$ is the number of peptide bonds in the
molecule, $R$ the ideal gas constant and $T$ the absolute
temperature. $m_k \in \{ 0, 1 \}$ is the binary variable which tells
whether the $k$--th bond, i.e. the one between residues $k$ and $k+1$, 
is in the disordered (0) or ordered (1)
state, and $q_k$ is the corresponding ordering entropic cost. The
product $
{\prod_{k=i}^j} m_k$ takes value 1 if and only
if all the peptide bonds from $i$ to $j$ are in the native state,
thereby realizing the assumed interaction.
The contact matrix with
elements $\Delta_{i,j} \in \{ 0, 1 \}$ tells which peptide bonds are at close distance in the native state; non--native interactions are disregarded. The contact map beween peptide bonds, $\Delta_{i,j}$ is derived from the standard one between residues, $\Delta_{i,j}^r$, as $\Delta_{i,j}=\Delta_{i,j+1}^r$, thus assigning the (residue $i$)-(residue $j$) contact to peptide bonds $i$ and $(j-1)$. 
$\Delta_{i,j}^r$ is usually calculated
according to some cutoff distance residues $i$ and $j$ in the native state; here we deal with ideal secondary structures, and will impose $\Delta_{i,j}^r$ accordingly.
Contact energy will be taken as $\epsilon_{i,j} = - \epsilon$
throughout the paper, unless differently stated. 
Without loss of generality, we can set $\epsilon =1$. 
Also the entropic costs will be taken to be homogeneous, $q_k = q = 2 \ln
2$. Several values of these entropies have been considered in various
works up to now, typically based on some fits to experimental
data. Here we want to consider model structures, so the specific value is not too relevant: we take $q = 2
\ln 2$, comparable to the results obtained for
various molecules; in the following 
we will discuss the $q$-dependence of our results.
Notice that, once fixed $\epsilon$ and $q$, the only source of heterogeneity comes from the contact matrix, i.e., from the geometry of the native state.

We define the denaturation temperature $T_m$ asking that  
the average fraction $m$ of
ordered bonds is halfway between its values $m_0 = 1$ (at T=0) and $m_\infty =
\frac{1}{N}
\sum_{k=1}^N (1 + e^{q_k})^{-1}$
at infinite temperature. In the following, for each structure considered, we will always work at the corresponding $T_m$. 
Exact evaluation \cite{WS1,WS2} of thermodynamic quantities will be performed as in \cite{BP,P}.

The kinetic evolution of the model 
is
described through a
discrete--time master equation, $p_{t+1}(m) = \sum_{m'} W(m' \to m) p_t(m)$, for the probability distribution
$p_t(m)$ at time $t$, where $m = \{ m_k, \,k = 1, \ldots N \}$ denotes
the state of the system.
The transition matrix $W$ is specified by a single bond flip
Metropolis rule,
 as 
in \cite{ZP-PRL,ZP-JSTAT}.
The kinetics will be studied by means of the local equilibrium
approach \cite{ZP-PRL,ZP-JSTAT}, where the equilibration rate $k$ can
be computed as the largest eigenvalue of a matrix of rank
$N(N+1)/2$. It has been shown in \cite{ZP-PRL,ZP-JSTAT} that this
approach turns out to be very accurate when compared to exact or Monte
Carlo results, and 
the rate so obtained is an upper bound of the
exact one. Notably, this approach allows us to evaluate directly relaxation rates, which are the experimentally accessible quantities (at difference with folding and unfolding rates), 
without choosing 
a reaction coordinate.

We apply the model first to parallel structures (see Supplementary Material, Fig.~1), where
the sequence distance of any interacting pair of residues is constant. This
class includes 
$\alpha$--helices and parallel
$\beta$--sheets. The first structural indicator we
consider is the ACO, defined as 
\begin{equation}
{\rm ACO} = N_c^{-1} \displaystyle{\sum_{1 \le i < j \le N}} \Delta_{i,j}
(j-i+1),
\end{equation}
where $N_c =\sum_{1 \le i < j \le N} \Delta_{i,j}$ is the total number of contacts, and 
we add 1 to the usual
$j-i$, since our contact matrix is defined
with reference to peptide bonds, and the number of the latter
involved in a contact is the corresponding number of residues minus
one.

\begin{figure}
\centerline{\includegraphics[width=0.4 \textwidth 
]{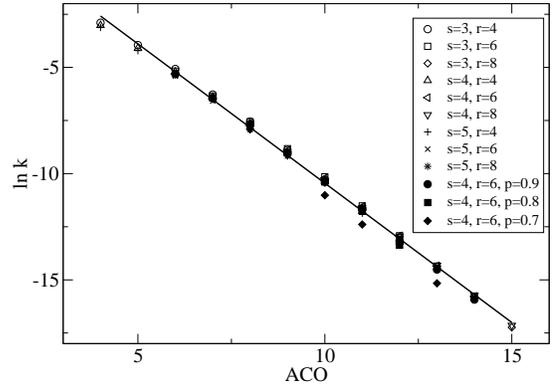}}
\caption{\label{fig:para} Logarithm of equilibration rate vs.\
  absolute contact order for parallel structures. Symbols explained in
  text.}
\end{figure}

In Fig.\ \ref{fig:para} we report the natural logarithm of the
equilibration rate at 
$T_m$ for several
parallel structures, 
defined as parallel $\beta$--sheets
with $s$ strands and $r$ residues per strand, where consecutive
strands are separated by loops of $l$ residues 
not involved in any 
contact. In such a structure the number of peptide bonds is $N = s
(r+l) - l - 1$ and ACO $ = r+l$; contact matrix elements are $\Delta_{i,j}=1$ if  
$j= i+l+r-1$ and $\Delta_{i,j}=0$  otherwise. The $\alpha$--helix corresponds to
the case $l=0$, ACO $ = r=4$. For every $(s,r)$ pair we consider values
of $l$ from 0 up to 8 in order to vary the ACO. The effect of dilution
is also considered (for $s=4$, $r=6$ only), where this regular
structure is perturbed by removing contacts with probability
$1-p$. The value of $p$, when not otherwise specified, 
is 1.

We see that all the results fall almost perfectly on the same
straight line. A joint linear fit including all data yields $\ln k =
2.6633 - 1.3113$ ACO, with correlation coefficient -0.998. Considering
different values for the entropic costs would not change the overall
behaviour but only the slope. For instance, $q = 2$ yields a slope
-1.73.

This result means that, according to this model, the ACO is
of fundamental importance, while the rate cannot depend in a relevant way on other
measures like the CO or the chain length, since each ACO value corresponds to several values of CO and total length. LRO, total contact distance,
the fraction of short range contacts, and heterogeneity in contact
distance are not applicable here, 
since all contacts present the same sequence distance.
Even the introduction of contact-energy heterogeneities does not affects the rate:
considering $s=4$, $r=7$ and $l=0$ and 4, and taking  $\epsilon_{i,j}$ uniformly distributed in $[-\epsilon - \Delta
\epsilon, -\epsilon + \Delta \epsilon]$, with $\Delta \epsilon \in [0,\epsilon ]$, we found that the variation of $\ln k$ was compatible
with the spread in Fig.\ \ref{fig:para}: less than 0.02 in absolute
value for $l=0$ ($\ln k \simeq -5.45$ in the uniform case), and less
than 0.6 for $l=4$ ($\ln k \simeq -10.39$ in the uniform case).
These figures hardly change if we consider the interactions as product of charges associated to residues, as in \cite{Tiana2004}, and take random charges, or if we consider a correlated disorder, e.g. strenghtening the contacts of a group of three consecutive residues, leaving the others unchanged.
In view of the above result it is worth checking whether this almost
perfect linear behaviour still holds  for structures which
maintain some regularity, but where the contact distance is not
constant. 
We therefore turn to antiparallel structures, 
and extend the results reported in \cite{ZP-PRL} for a four--stranded antiparallel
$\beta$--sheet,
considering 
sheets with
$s$ strands and $r$ residues per strand; in the case $s = 2$ we
have a simple hairpin (see Supplementary Material, Fig.~2). 
Turns corresponds to peptide bonds $k r$; contacts are established between bonds $i$ and $j=2 k r -i$, with $(k-1)r+1 \leq i \leq  kr-1$ and $k=1,\cdots,s-1$ labelling the turns; the total number of peptide bonds is $N = s r - 1$. 
This represents the most connected case: every residue but two (in the first and last turn) has at least one contact. The contact distance varies from 3 to $2r-1$ in steps of 2, and hence ACO $ = r+1$. 

\begin{figure}
\centerline{\includegraphics[width=0.45 \textwidth]{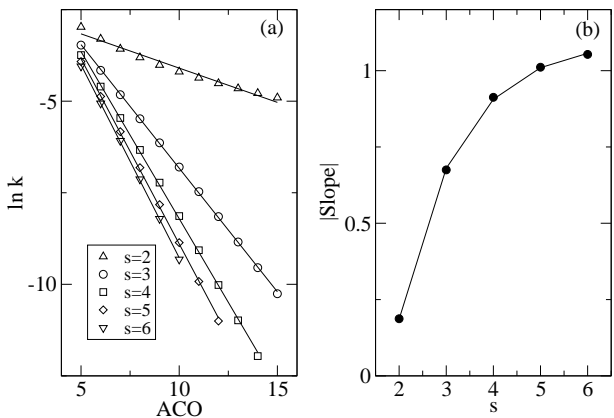}}
\caption{\label{fig:anti} (a) Logarithm of equilibration rate vs.\
  absolute contact order for antiparallel structures at the denaturation temperature $T_m$. Symbols
  explained in text. (b) Absolute value of the slope in (a) as a
  function of $s$.}
\end{figure}

In Fig.\ \ref{fig:anti}(a) we report the natural logarithm of the
equilibration rate at 
$T_m$ as a function of
ACO for several $(s,r)$ values. It is clearly seen that the almost
perfect linear correlation is now valid only at fixed strand number
$s$ (again, slopes would be larger for a larger entropic cost
$q$). The absolute value of the slope increases with $s$ and appears
to tend to a constant for large $s$. Indeed, a fit to an exponential
function is made in panel (b) and is almost perfect 
($\vert {\rm Slope} \vert = 1.21 -3.5 \exp(-0.62 s)$, estimated variance is 0.00017, correlation
coefficient is 0.9996). 

The limiting value of the slope is slightly smaller
than 
that
obtained for parallel structures,
whence
we have that, at fixed ACO$\gtrsim 4$, the hairpin is the fastest structure, then
the other antiparallel structures come, in order of increasing $s$,
and finally we have the parallel structures.

The exponential dependence of the logarithm of the rate on $s$ is quite puzzling: to understand such a behavior, we study the folding mechanism, as revealed by the probabilities of native "strings" $S_{i,j}$, i.e., of configurations where peptide bonds $i$ and $j$ are unfolded, while all those in between them are native. Assuming that the relaxation rate can be related to the height of the highest barrier along the folding pathway, the analysis reveals that the latter is represented by the folding of one hairpin, or more precisely of all the residues between peptide bond $u=(k-1) r$ and $v=(k+1) r+1$, being $u$ and $v$ unfolded, for any $k$ between 1 and $(s-1)$. 
The probability of such a configuration can be estimated as $p(S_{u,v})=\exp(\beta_m \epsilon (r-1)-2 r q)/Z_{u,v}$, where $\beta_m=1/(R T_m)$. $Z_{u,v}$ is the partition function restricted to the region between $u$ and $v$. It can be estimated as
\begin{eqnarray}
Z_{u,v} &=&A^{2 r -1}+A e^{\beta_m \epsilon (r-1)-(2 r-1)q}  \nonumber \\
&+& (2 e^{-q}+1) \sum_{j=0}^{r-2} A^{2(r -j)-3} e^{\beta_m \epsilon j-(2 j+1)q} ,
\end{eqnarray}
where 
$A=(1+\exp(-q))$. 
If our system can be considered a two-state folder we have that the relaxation rate 
$k=k_f+k_u$ is the sum of the folding and unfolding rates; the latter will be identical at $T=T_m$ and, 
assuming an Arrhenius scheme, they will be proportional to the folding probability   $p(S_{u,v})$. So, if the above assumptions hold and we have identified the correct barrier, we will find the same behavior for $\ln p(S_{u,v})$  as for $\ln k$.
Notice that the expression of $p(S_{u,v})$ contains clearly the ACO, since $r$=ACO-1, while the only quantity depending on $s$ is the value of $\beta_m$: one finds indeed that the exponential dependence  of $\ln k$ on $s$ is related 
to the dependence of $\beta_m$ on s and ACO.
Taking this into account in the expression of $p(S_{u,v})$ yields a dependence on ACO which is completely analogous to that reported in Fig.~
\ref{fig:anti}, where again  $\ln p(S_{u,v})$ can be fitted by straight lines whose slopes follow the law: 
$\vert {\rm Slope} \vert =  1.40 - 5.1 \exp(- 0.70 s )$, (estimated variance: 0.0012, correlation coefficient: 0.9980)
reasonably close to the value obtained for the rates.

The above result is very interesting, because 
it proves that an Arrhenius framework can still be applied in this case, despite the 
free-energy profiles may present more than one minimum and barrier. But even more, these results establish a quantitative connection between the rate and the number of strands 
through the stability, stating that as the number of strands of the system increases, the $T_m$ increases asymptotically, 
implying a global increase of the stability, while the equilibration rate decreases.
Notice that the dependence of the rate on $s$ found in the model is consistent with the reported observation of a  $L^\alpha$ - dependence \cite{Ivankov}: indeed if we fit the rate to the number of residues $L=s r$ we find for the antiparallel structures 
$\ln k= -0.73  \times L^{0.64}$, which is 
not far from the experimental exponent reported in \cite{ Ivankov} ($\alpha =0.70$) and from  $\alpha=0.61$ coming from off-lattice simulations \cite{Koga}. 
However, an accurate comparison of the dependence of rates on length should take into account the frequency of the different structures in real proteins, which is out of the scope of the present work.

To conclude, let us review the main findings:
resorting to a simple statistical model, we have performed a detailed analysis of the dependence of the relaxation rate on some structural indicators, known to correlate  
with protein folding times. 
To elucidate the role and the interplay of the different factors, we have studied ideal helical, parallel and antiparallel secondary structures, of different length and number of strands. 
Our results confirm the absolute contact order ACO as the main structural determinant of the rates, but suggest different folding mechanisms for parallel and antiparallel structures: the latter, which fold faster than the former at equal ACO, show a dependence on the number of strands which we can 
relate 
to the dependence of the denaturation temperature on the ACO and number of strands.
We also find that the dependence of the rate on the length of the protein is consistent with the power-law reported in \cite{Ivankov} for real proteins, despite the fact our ideal structures lack the structural and energetic heterogeneity of the latter. 

P.~B. acknowledges support from Spanish 
Education and Science 
Ministry  
(FIS2004-05073, FIS2006-12781).


\end{document}